\documentclass[10pt]{article}
\usepackage{geometry} 
\usepackage[hang,bf]{caption}
\usepackage{setspace}
\usepackage[parfill]{parskip}    
\usepackage{times}
\usepackage{graphicx}
\usepackage{amssymb,amsmath,amsbsy}
\usepackage{cancel}
\usepackage{epstopdf}
\usepackage{threeparttable}
\usepackage{epsfig}
\usepackage{cite}
\newcommand{\nbb}{0\nu\beta\beta}
\newcommand{\bb}{\beta\beta}

\begin{document}
\title{\textbf{Results from a search for the $\pmb{\nbb}$-decay of $\mathbf{^{130}Te}$\\\small{(Physical Review C $\mathbf{78}$, 035502 (2008))}}}
\author{C. Arnaboldi$^{1,2}$, D.R. Artusa$^3$, F.T. Avignone III$^{3}$\footnote{Corresponding author}, M. Balata$^4$, I.C. Bandac$^3$, M. Barucci$^{5,6}$, \\ J.W. Beeman$^7$, F. Bellini$^{17,18}$, C. Brofferio$^{1,2}$, C. Bucci$^4$, S. Capelli$^{1,2}$, L. Carbone$^2$, \\ S. Cebrian$^8$, M. Clemenza$^{1,2}$, O. Cremonesi$^2$, R.J. Creswick$^3$, A. de Waard$^9$,\\ S. Di Domizio$^{10,11}$, M. J. Dolinski$^{14,15}$, H.A. Farach$^3$, E. Fiorini$^{1,2}$, G. Frossati$^9$, A. Giachero$^4$, \\ A. Giuliani$^{12,2}$, P. Gorla$^4$, E. Guardincerri$^7$, T. D. Gutierrez$^{19}$, E.E. Haller$^{7,13}$,\\ R.H. Maruyama$^{20}$, R.J. McDonald$^7$,  S. Nisi$^4$, C.  Nones$^{12,2}$, E.B. Norman$^{14,21}$, A. Nucciotti$^{1,2}$,\\ E.  Olivieri$^{5,6}$, M. Pallavicini$^{10,11}$, E. Palmieri$^{16}$, E. Pasca$^{5,6}$, M. Pavan$^{1,2}$, M. Pedretti$^2$, G. Pessina$^2$, \\ S. Pirro$^2$, E. Previtali$^2$, L. Risegari$^{5,6}$, C. Rosenfeld$^3$, S. Sangiorgio$^{20}$, M. Sisti$^{1,2}$,\\ A.R. Smith$^7$, L. Torres$^8$, G. Ventura$^{5,6}$, M. Vignati$^{17,18}$.}
\date{}
\maketitle
\vspace{-0.45in}

\begin{center}
\small{$^1$ Dipartimento di Fisica dell'Universit\`a di Milano-Bicocca, I-20126 Milano, Italy\\
$^2$ Sesione INFN di Milano-Bicocca, I-20126 Milano, Italy\\
$^3$ Department of Physics and Astronomy, University of South Carolina, Columbia, SC 29208, USA\\
$^4$ INFN Laboratori Nazionali del Gran Sasso, I-67010, Assergi (L'Aquila), Italy\\
$^5$ Dipartimento di Fisica dell'Universit\`{a} di Firenze, I-50019 Firenze, Italy\\ 
$^6$ Sezione INFN di Firenze, I-50019 Firenze, Italy\\
$^7$ Lawrence Berkeley National Laboratory, Berkeley, CA 94720, USA\\ 
$^8$ Laboratorio de Fisica Nuclear y Altas Energias, Universidad de Zaragoza, E-50001 Zaragoza, Spain\\
$^9$ Kamerling Onnes Laboratory, Leiden University, 2300 RAQ Leiden, Netherlands\\
$^{10}$ Dipartimento di Fisica dell'Universit\`{a} di Genova, I-16146 Genova, Italy\\
$^{11}$ Sezione INFN di Genova, I-16146 Genova, Italy\\
$^{12}$ Dipartimento di Fisica e Matematica dell'Universit\`{a} dell'Insubria, I-22100 Como, Italy\\
$^{13}$ Department of Materials Science and Engineering, University of California, Berkeley, CA 94720, USA\\
$^{14}$ Lawrence Livermore National Laboratory, Livermore, CA 94551, USA\\
$^{15}$ Department of Physics, University of California, Berkeley, CA 94720, USA\\
$^{16}$ INFN Laboratori Nazionali di Legnaro, I-35020 Legnaro (Padova), Italy\\
$^{17}$ Dipartimento di Fisica dell'Universit\`{a} di Roma La Sapienza, I-00185 Roma, Italy\\
$^{18}$ Sezione INFN di Roma, I-00185 Roma, Italy\\
$^{19}$ Physics Department, California Polytechnic State University, San Luis Obisbo, CA 93407 USA\\
$^{20}$ Department of Physics, University of Wisconsin, Madison, WI 53706, USA\\
$^{21}$ Department of Nuclear Engineering, University of California, Berkeley, CA 94720, USA}
\end{center}

\begin{abstract}\
A detailed description of the CUORICINO $^{130}Te$ neutrinoless double-beta ($\nbb$) decay experiment is given and recent results are reported. CUORICINO is an array of $62$ tellurium oxide ($TeO_{2}$) bolometers with an active mass of $40.7$ kg. It is cooled to $\sim 8-10$ mK by a dilution refrigerator shielded from environmental radioactivity and energetic neutrons. It is running in the Laboratori Nazionali del Gran Sasso (LNGS) in Assergi, Italy. These data represent an exposure of  $11.83\textrm{ kg}\cdot\textrm{y}$ or $91$ mole-years of $^{130}Te$. No evidence for $\nbb$-decay was observed and a limit of $T^{0\nu}_{1/2}\left(^{130}Te\right)\geq3.0\times10^{24}$ y ($90\%$ C.L.) is set. This corresponds to an upper limit on the effective mass, $\left\langle m_{\nu}\right\rangle$, between $0.19$ and $0.68$ eV when analyzed with the many published nuclear structure calculations. In the context of these nuclear models, the values fall within the range corresponding to the claim of evidence of $\nbb$-decay by H.V. Klapdor-Kleingrothaus, \textit{et al.} The experiment continues to acquire data.
\end{abstract}


\section{INTRODUCTION}\label{sec:1}
There are three very important open questions in neutrino physics that can best be addressed by next generation neutrinoless double-beta $\left(\nbb\right)$ decay experiments.  First, are neutrinos Majorana particles that differ from antineutrinos only by helicity? Second, what is their mass-scale? Third, is lepton number conservation violated? While searches for  $\bb$-decay have been carried out steadily throughout many decades \cite{1,2,3}, it is now a far more interesting time for the field. Atmospheric neutrino-oscillation data imply that there exist scenarios in which the effective Majorana mass of the electron neutrino could be larger than $0.05$ eV. Recent developments in detector technology make the observation of $\nbb$-decay at this scale now feasible. For recent comprehensive experimental and theoretical reviews see \cite{4,5,6}. Optimism that a direct observation of $\nbb$-decay is possible was greatly enhanced by the observation and measurement of the oscillations of atmospheric neutrinos \cite{7}, the confirmation by SuperKamiokande \cite{8} of the deficit of $^{8}B$ neutrinos observed by the chlorine experiment \cite{9}, the observed deficit of $p-p$ neutrinos by SAGE \cite{10} and GALEX \cite{11}, and the results of the SNO experiment \cite{12} that clearly showed that the total flux of $^{8}B$ neutrinos from the sun predicted by Bahcall and his co-workers \cite{13} is correct. Finally, the data from the KamLAND reactor-neutrino experiment strongly favor the MSW large mixing-angle solution of solar neutrino oscillations \cite{14}. This important list of results published since $1998$ weighs very heavily in favor of supporting two or more next generation $\nbb$-decay experiments (see the reports in references \cite{15,16}).

\begin{sloppypar}
The most sensitive limits have come from germanium detectors enriched in $^{76}Ge$. They were the Heidelberg-Moscow experiment $\left(T_{1/2}^{0\nu}\left(^{76}Ge\right)\geq1.9\times10^{25}y\right)$ \cite{17} and the IGEX experiment $\left(T_{1/2}^{0\nu}\left(^{76}Ge\right)\geq1.6\times10^{25}y\right)$ \cite{18}. These imply that the upper bound on the effective Majorana mass of the electron neutrino, $\langle m_{\nu}\rangle$, defined below, ranges from $\sim 0.3$ to $\sim 1.0$ eV, depending on the choice of nuclear matrix elements used in the analysis. However, a subset of the Heidelberg-Moscow Collaboration has reanalyzed the data and claimed evidence of a peak at the total decay energy, $2039$ keV, implying $\nbb$-decay \cite{19,20}. While there have been opposing views \cite{21,22,23}, there is no clear proof that the observed peak is not an indication of $\nbb$-decay. The GERDA experiment, also using $^{76}Ge$, is under construction in the Laboratori Nazionali del Gran Sasso (LNGS), and will test this claim \cite{24}. The CUORICINO experiment, also located at LNGS, is the most sensitive $\nbb$-decay experiment with good energy resolution currently operating \cite{25,26}. It is searching for the $\nbb$-decay of $^{130}Te$ and has the capability of confirming the claim; however, a null result cannot be used to refute the claim because of the uncertainty in the nuclear matrix element calculations. The proposed Majorana $^{76}Ge$ experiment \cite{27}, CUORE $^{130}Te$ experiment \cite{28}, and EXO $^{136}Xe$ experiment \cite{29} are all designed to reach the $\langle m_{\nu}\rangle\approx0.05$ eV mass sensitivity and below. Descriptions of other proposed experiments with similar goals are given in the recent reviews \cite{4,5,6}.
\end{sloppypar}

There are other constraints on the neutrino-mass scale, irrespective of their Majorana or Dirac character. The Troitsk \cite{30} and Mainz \cite{31} $^{3}H$ single $\beta$-decay experiments have placed an upper limit of $2.2$ eV on the mass of the electron neutrino. The KATRIN experiment, a greatly enlarged $^{3}H$ $\beta$-decay experiment in preparation, is projected to have a sensitivity of $0.2$ eV \cite{32}.

Astrophysical data are also very relevant in a discussion of neutrino mass. In a recent paper by Barger \emph{et al.}, \cite{33} an upper limit on the sum of neutrino mass eigenvalues, $\Sigma\equiv m_{1}+m_{2}+m_{3}\leq 0.75$ eV ($95\%$ C.L.), was derived. The data used were from the Sloan Digital Sky Survey (SDSS) \cite{34}, the two degree Field Galaxy Red Shift Survey (2dFGRS) \cite{35}, and the Wilkinson Microwave Anisotropy Probe (WMAP) \cite{36}, as well as other CMB experiments and data from the Hubble Space Telescope. Hannestad \cite{37} used the WMAP and 2dFGRS data to derive the bound $\Sigma<1.0$ eV ($95\%$ C.L.) and concluded that these data alone could not rule out the evidence claimed in \cite{19,20}. On the other hand, Allen, Schmidt and Briddle \cite{38} found a preference for a non-zero neutrino mass, i.e., $\Sigma=0.56^{+0.30}_{-0.25}$ eV. This is interestingly close to the favored range of values given in \cite{19,20}. For recent papers on the subject see \cite{39} and references therein. The constraint $\Sigma\leq 0.75$ eV would imply that the lightest neutrino eigenstate mass $m_{1}<0.25$ eV. On the other hand, if the claim of the positive value of $\Sigma$ would be  correct, $\left\langle m_{\nu}\right\rangle\approx 0.17$ eV, and next generation $\nbb$-decay experiments would constitute a stringent test of lepton-number conservation, irrespective of the neutrino mass hierarchy (see the discussion of hierarchy below).

In this paper we present a detailed description and present the results from the CUORICINO $\nbb$-decay experiment derived from data taken between April $2003$ and May $2006$. Finally, we note that $^{130}Te$ has a series of calculated matrix elements implying values of $\left\langle m_{\nu}\right\rangle$ derived from the CUORICINO half-life limit between $\sim 0.20$ keV, and $\sim 0.68$ keV. A detailed discussion of the implications from the recent developments in the theoretical nuclear structure calculations is given later.


\section{NEUTRINO PHYSICS AND NEUTRINOLESS\\ DOUBLE-BETA DECAY}\label{sec:2}
Neutrino-oscillation data very strongly imply that there are three neutrino flavor eigenstates, $\left|\nu_{e,\mu,\tau}\right\rangle$, that are super positions of three mass eigenstates, $\left|\nu_{1,2,3}\right\rangle$, of the weak Hamiltonian as expressed in equation \eqref{eq:1}:
 
 \begin{equation}
 \label{eq:1}
 \left|\nu_{l}\right\rangle=\sum^{3}_{j=1}\left|u_{lj}^{L}\right|e^{i\delta_{j}}\left|\nu_{j}\right\rangle,
 \end{equation}
 
where $l=e,\mu,\tau$, and the factor $e^{i\delta_{j}}$ is a CP phase, $\pm1$ for CP conservation.
 
The decay rate for the $\nbb$-decay mode driven by the exchange of a massive Majorana neutrino is expressed in the following approximation:
 
 \begin{equation}
 \label{eq:2}
 \left(T^{0\nu}_{1/2}\right)^{-1}=G^{0\nu}\left(E_{0},Z\right)\left|\frac{\left\langle m_{\nu}\right\rangle}{m_{e}}\right|^{2}\left|M_{f}^{0\nu}-\left(g_{A}/g_{V}\right)^{2}M_{GT}^{0\nu}\right|^{2},
 \end{equation}

where $G^{0\nu}$ is a phase space factor including the couplings, $\left|\left\langle m_{\nu}\right\rangle\right|$ is the effective Majorana mass of the electron neutrino discussed below, $M_{f}^{0\nu}$ and $M_{GT}^{0\nu}$ are the Fermi and Gamow-Teller nuclear matrix elements respectively, and $g_{A}$  and $g_{V}$ are the relative axial-vector and vector weak coupling constants respectively. After multiplication by a diagonal matrix of Majorana phases, $\left\langle m_{\nu}\right\rangle$ is expressed in terms of the first row of the $3\times 3$ matrix of equation \eqref{eq:1} as follows: 

\begin{equation}
\label{eq:3}
\left|\left\langle m_{\nu}\right\rangle\right|\equiv\left|\left(u_{e1}^{L}\right)^{2}m_{1}+\left(u_{e2}^{L}\right)^{2}m_{2}e^{i\phi_{2}}+\left(u_{e3}^{L}\right)^{2}m_{3}e^{i\left(\phi_{3}+\delta\right)}\right|,
\end{equation}

where $e^{i\phi_{2,3}}$ are the Majorana CP phases ($\pm 1$ for CP conservation in the lepton sector). Only the phase angle $\delta$ appears in oscillation
 expressions. The two Majorana phases, $e^{i\phi_{2,3}}$, do not, and hence do not affect neutrino oscillation measurements. The oscillation experiments have, however, constrained the mixing angles and thereby the coefficients $u_{lj}^{L}$ in equation \eqref{eq:3}. Using the best-fit values from the SNO and Super Kamiokande solar neutrino experiments and the CHOOZ \cite{40}, Palo Verde \cite{41} and KamLAND \cite{14} reactor neutrino experiments, we arrive at the following expression in the case of the normal hierarchy:

\begin{equation}
\label{eq:4}
\left|\left\langle m_{\nu}\right\rangle\right|=\left|\left(0.70^{+0.02}_{-0.04}\right)m_{1}+\left(0.30^{+0.04}_{-0.02}\right)m_{2}e^{i\phi_{2}}+\left(\leq 0.05\right)m_{3}e^{i\left(\phi_{3}+\delta\right)}\right|,
\end{equation}

where the errors are approximated from the published confidence levels (C.L.). The bound on $\left|u_{e3}\right|^{2}$ is at the $2\sigma$ C.L. and the errors on the first two coefficients are $1\sigma$. In the convention used here, the expression for the inverted hierarchy, discussed below, is obtained by exchanging $m_{1}\Leftrightarrow m_{3}$ in equation \eqref{eq:4}.

The results of the solar neutrino and atmospheric neutrino experiments yield the mass square differences $\delta_{ij}^{2}=\left|m_{i}^{2}-m_{j}^{2}\right|$ but cannot distinguish between two mass patterns (hierarchies): the "normal" hierarchy, in which $\delta m_{\textrm{solar}}^{2}=m_{2}^{2}-m_{1}^{2}$ and $m_{1}\cong m_{2}\ll m_{3}$, and the "inverted" hierarchy where  $\delta m_{\textrm{solar}}^{2}=m_{3}^{2}-m_{2}^{2}$ and $m_{3}\cong m_{2}\gg m_{1}$. In both cases we can approximate $\delta m_{AT}^{2}\cong m_{3}^{2}-m_{1}^{2}$ . Considering the values in equation \eqref{eq:4}, we make the simplifying approximation $\left(u_{e3}\right)^{2}\approx 0$. Using the central values of equation \eqref{eq:4}, we can write the following approximate expressions: 

\begin{equation}
\label{eq:5}
\left|\left\langle m_{\nu}\right\rangle\right|\cong m_{1}\left|0.7+0.3e^{i\phi_{2}}\sqrt{1+\frac{\delta_{solar}^{2}}{m_{1}^{2}}}\right|,
\end{equation}

for the case of "normal"  hierarchy, and,

\begin{equation}
\label{eq:6}
\left|\left\langle m_{\nu}\right\rangle\right|\cong \sqrt{m_{1}^{2}+\delta m_{AT}^{2}}\left|0.7+0.3e^{i\phi_{2}}\right|,
\end{equation}

in the "inverted" hierarchy case. At this time there is no experimental evidence favoring either hierarchy. In Table \ref{tab:1}, we use Eqs. \eqref{eq:5} and \eqref{eq:6} to show the predicted central values of $\left\langle m_{\nu}\right\rangle$ as a function of the lightest neutrino mass eigenvalue, $m_1$. These values roughly define the desired target sensitivities of next generation $\nbb$-decay experiments.

\begin{table}[htdp]
\caption{Central values of the numerical predictions of $\left|\left\langle m_{\nu}\right\rangle\right|$ (meV) for both hierarchies and CP phase relations. ( $m_{1}$ is also given in meV.)}
\begin{center}
\vspace{5pt}
\begin{tabular}{ccccccccccc}
\hline\hline
\\
\multicolumn{5}{c}{Normal Hierarchy} & &\multicolumn{5}{c}{Inverted Hierarchy} \\
\\
\cline{1-5}\cline{7-11}
\multicolumn{2}{c}{$e^{i\phi_{2}}=-1$} && \multicolumn{2}{c}{$e^{i\phi_{2}}=+1$} & &\multicolumn{2}{c}{$e^{i\phi_{2}}=-1$} && \multicolumn{2}{c}{$e^{i\phi_{2}}=+1$} \\
\cline{1-2}\cline{4-5}\cline{7-8}\cline{10-11}
$m_{1}$ & $\left|\left\langle m_{\nu}\right\rangle\right|$ && $m_{1}$ & $\left|\left\langle m_{\nu}\right\rangle\right|$ & &$m_{1}$ & $\left|\left\langle m_{\nu}\right\rangle\right|$ && $m_{1}$ & $\left|\left\langle m_{\nu}\right\rangle\right|$ \\
\hline
$20.0$&$7.90$&&$20.0$&$20.2$&&$0.00$&$20.0$&&$0.00$&$50.0$ \\
$40.0$&$16.0$&&$40.0$&$40.0$&&$20.0$&$21.6$&&$20.0$&$53.9$\\
$60.0$&$24.0$&&$60.0$&$60.0$&&$50.0$&$28.3$&&$50.0$&$70.7$\\
$80.0$&$32.0$&&$80.0$&$80.0$&&$75.0$&$36.0$&&$75.0$&$90.1$\\
$100.0$&$40.0$&&$100.0$&$100.0$&&$100.0$&$44.7$&&$100.0$&$111.0$ \\
$200.0$&$80.0$&&$200.0$&$200.0$&&$200.0$&$82.5$&&$200.0$&$206.0$ \\
$400.0$&$160.0$&&$400.0$&$400.0$&&$400.0$&$161.1$&&$400.0$&$403.0$ \\
\hline\hline
\end{tabular}
\end{center}
\label{tab:1}
\end{table}

It is clear that a next generation experiment should have at least the sensitivity for discovery in the case of an inverted hierarchy when $e^{i\phi_{2}}=e^{i\phi_{3}}$ and for $m_{1}=0$. In this case, $\left\langle m_{\nu}\right\rangle\approx\sqrt{\delta_{\textrm{AT}}^2}\approx 0.050\textrm{ eV}$. It should also be capable of being expanded in case this level is reached and no effect is found \cite{15,16}.

It is convenient to define the nuclear structure factor, $F_{N}$, (sometimes denoted as $C_{mm}$ in the literature) as follows:

\begin{equation}
\label{eq:7}
F_{N}\equiv G^{0\nu}\left|M_{f}^{0\nu}-\left(g_{A}/g_{V}\right)^{2}M_{GT}^{0\nu}\right|^{2}.
\end{equation}

Accordingly, the effective Majorana mass of the electron neutrino is connected to the half-life as shown in equation \eqref{eq:8}:

\begin{equation}
\label{eq:8}
\left\langle m_{\nu}\right\rangle=\frac{m_{e}}{\sqrt{F_{N}T_{1/2}^{0\nu}}}.
\end{equation}

To extract values of $F_N$ from theoretical papers, we recommend using their calculated values of half lives for a given value of $\left\langle m_\nu\right\rangle$, thereby avoiding difficulties associated with conventions used in calculating phase-space factors.

Possible interpretations of the null result of CUORICINO, in terms of the effective Majorana neutrino mass, may be understood with detailed analyses of the nuclear matrix elements discussed in a Secs. \ref{sec:8} and \ref{sec:9}. In Sec. \ref{sec:10}, this null result will be compared with the positive claim report reported in \cite{19,20}.

\section{THE EXPERIMENT}\label{sec:3}

The CUORICINO experiment is an array of cryogenic bolometers containing $^{130}Te$, the parent $\nbb$-decay isotope. This technique was suggested for $\beta\beta$-decay searches by Fiorini and Niinikoski \cite{42} and applied earlier by the Milano group in the MIBETA experiment \cite{43}. The bolometers are sensitive calorimeters that measure the energy deposited by particle or photon interactions by measuring the corresponding rise in temperature. The CUORICINO bolometers are single crystals of $TeO_{2}$; they are dielectric and diamagnetic, and are operated at temperatures between $8$ and $10$ mK \cite{44,45}. According to the Debye Law, the specific heat of $TeO_{2}$ crystals is given by $C(T)=\beta\left(T/\Theta_{D}\right)^{3}$, where $\beta=1994\textrm{ JK}^{-1}\textrm{mol}^{-1}$ and $\Theta_{D}$ is the Debye temperature. In these materials, $C(T)$ is due almost exclusively to lattice degrees of freedom. A special measurement determined the value of  $\Theta_{D}$, as $232$ K \cite{43}. This differs from the previously published value of $272$ K \cite{46}. The specific heat followed the Debye Law down to $60$ mK. The heat capacity of these crystals, extrapolated to $10$ mK, is $2.3\times10^{-9}\textrm{ JK}^{-1}$. With these values of the parameters, an energy deposition of a few keV will result in a measurable temperature increase, $\Delta T$. In CUORICINO, $\Delta T$ is measured by high-resistance germanium thermistors glued to each crystal. More details can be found in reference \cite{44} and in earlier publications \cite{47,48}. Accordingly, the temperature increase caused by the deposition of energy equal to the total $\beta\beta$-decay energy, $Q_{\beta\beta}=2530.3\pm 2.0$ keV, would be $1.77\times10^{-4}$ K. To obtain usable signals for such small temperature changes, very sensitive thermistors are required.

The thermistors are heavily doped high-resistance germanium semiconductors with an impurity concentration slightly below the metal-insulator transition. High quality thermistors require a very homogeneous doping concentration. CUORICINO uses Neutron Transmutation Doped (NTD) germanium thermistors. This is achieved by means of uniform thermal neutron irradiation throughout the entire semiconductor volume, in a nuclear reactor. The electrical conductivity of these devices, which is due to variable range hopping (VHR) of the electrons, depends very sensitively on the temperature. The resistivity varies with temperature according to $\rho=\rho_{0}\exp{{\left(\frac{T_{0}}{T}\right)^{\gamma}}}$, where $\rho_{0}$ and $T_{0}$ depend on the doping concentration and $\gamma=1/2$. 

Thermistors can be parameterized by their sensitivity, $A(T)$, defined as follows: $A(T)\equiv\left|d\left(\ln{R}\right)/d\left(\ln{T}\right)\right|$ $=\gamma\left(T_{0}/T\right)^{\gamma}$, and where the resistance is $R(T)=R_{0}\exp{\left(T_{0}/T\right)^{\gamma}}$. The parameter $R_{0}\equiv\rho_{0}(d/a)$, where $d$ and $a$ are the distance between the contacts and the cross section of the thermistor, respectively. The values of $R_{0}, T_{0}$ and $\gamma$ were experimentally measured for about one third of the thermistors, and the average values used for the rest. The measurements were done by coupling the thermistor to a low-temperature heat sink with a high-heat-conductivity varnish glue, which can be easily removed with alcohol. The base temperature of the heat sink is between $15$ and $50$ mK \cite{50}. A current flows through the device and an I-V load curve is plotted. The curve becomes very non-linear due to the power dissipation, which causes the dynamic resistance, the slope of the I-V curve, to invert from positive to negative. The characterization, as discussed in Ref. \cite{51} is done on the thermistors directly mounted on a heat sink, while the optimum bias is studied for the complete detector, thermistor and crystal, since the noise figure depends on all thermal conductances, glue, wires, Teflon, etc. This allows the maximization of the signal to noise ratio. The parameters of each thermistor are determined from a combined fit to a set of load curves measured at different base temperatures. A detailed description of the characterization process for $Si$ thermistors was described in Ref. \cite{51} and same process was used for the CUORICINO $Ge$ thermistors.

The thermistors used in the MIBETA and CUORICINO experiments were specially developed and produced for this application \cite{52}. It is necessary to optimize the neutron doping of the $Ge$. This is facilitated by foils of metal with long-lived $(n, \gamma)$ radioactive daughter nuclides, allowing the neutron exposure to be evaluated without having to wait for the intense radiation of the $^{71}Ge$ in the $Ge$ sample to decay. Following the decay period, the $Ge$ is heat treated to repair the crystal structure and then cut into $3\times 3\times 1$ mm strips. Electrical connections are made with two $50\,\mu\textrm{m}$ gold wires, ball bonded to metalized surfaces on the thermistor. The thermistors are glued to each bolometer by nine spots of epoxy, deposited by an array of pins for better control of the thermal conductances and to minimize stresses at the interface between the two materials.


\section{THE CUORICINO DETECTOR}\label{sec:4}

\begin{figure}[htb]
\begin{center}
\epsfig{file=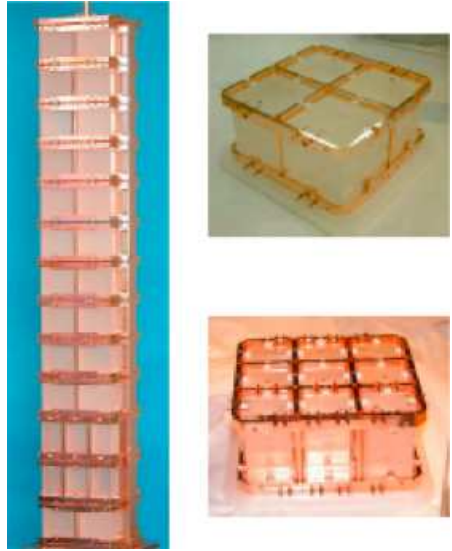, scale=0.6}
\caption{(Color online) The Tower of CUORICINO and individual $4$ and $9$ detector modules.}
\label{fig:1}
\end{center}
\end{figure}

CUORICINO is a pilot experiment for a larger experiment, CUORE (Cryogenic Underground Observatory for Rare Events) discussed later. It is a tower of $13$ planes \cite{25,26}. As shown in Fig. \ref{fig:1}, the CUORICINO structure is as follows: each of the upper $10$ planes and the lowest one consists of four $5\times 5\times 5\textrm{ cm}^{3}$ $TeO_{2}$ crystals (of natural isotopic abundance of $^{130}Te$) as shown in the upper right hand figure, while the $11^{th}$ and $12^{th}$ planes have nine, $3\times 3\times 6\textrm{ cm}^{3}$ crystals, as shown in the lower right hand figure. In the $3\times 3 \times 6\textrm{ cm}^{3}$ planes the central crystal is fully surrounded by the nearest neighbors for greater veto capability. 

The smaller crystals are of natural isotopic abundance except for four. Two of them are enriched to $82.3\%$ in $^{128}Te$ and two are enriched to $75\%$ in $^{130}Te$. All crystals were grown with pre-tested low radioactivity material by the Shanghai Institute of Ceramics and shipped to Italy by sea to minimize the activation by cosmic ray interactions. They were lapped with specially selected low contamination polishing compound. All these operations, as well as the mounting of the tower, were carried out in a nitrogen atmosphere glove box in a clean room. The mechanical structure is made of oxygen-free high-conductivity copper and Teflon, and both were previously tested to be sure that radioactive contaminations were minimal and consistent with the required detector sensitivity.

Thermal pulses are measured with NTD $Ge$ thermistors thermally coupled to each crystal. The thermistors are biased through two high-impedance load resistors at room temperature, with resistances typically in excess of one hundred times that of the thermistors. The large ratio of the resistances of the load resistors over those of the thermistors allows the parallel noise to be kept at an adequate level. Low frequency load-resistor noise was minimized by a specially designed circuit \cite{53}. The voltage signals from the thermistors are amplified and filtered before being fed to an analog-to-digital converter (ADC). This part of the electronic system is DC coupled, and only low-pass anti-aliasing filters are used to reduce the high-frequency noise. The typical bandwidth is approximately $10$ Hz, with signal rise and decay times of order $30$ and $500$ ms, respectively. This entire chain of electronics makes a negligible contribution to the detector energy resolution. More details of the design and features of the electronic system are found in \cite{54}. The gain of each bolometer is stabilized by means of a $Si$ resistor of $50$-$100$ $\textrm{k}\Omega$, attached to each bolometer that acts as a heater. Heat pulses are periodically supplied by a calibrated ultra-stable pulser \cite{55}. This sends a calibrated voltage pulse to the $Si$ resistor. This pulse has a time duration very much shorter than the typical thermal response of the detector \cite{44}. The Joule dissipation from the $Si$ resistor produces heat pulses in the crystal almost indistinguishable in characteristic shape from those from calibration $\gamma$-rays. The heater pulses are produced with a frequency of about one in every $300$ s in each of the CUORICINO bolometers. Any variation in the voltage amplitude recorded from the heater pulses indicates that the gain of that bolometer has changed. The heater pulses are used to measure  (and later correct offline) for the gain drifts. Two other pulses, one at lower and one at higher energies, are sent to the same resistors with much lower frequency. The former is used to monitor threshold stability, and the latter to check the effectiveness of the gain stability correction.

The tower is mechanically decoupled from the cryostat to avoid heating due to vibrations. The tower is connected through a $25$ mm copper bar to a steel spring fixed to the $50$ mK plate of the refrigerator. The temperature stabilization of the tower is made by means of a thermistor and a heater glued to it. An electronic channel is used for a feed back system \cite{56}. The entire setup is shielded with two layers of lead of $10$ cm minimum thickness each. The outer layer is made of common low radioactivity lead, while the inner layer is made of special lead with a measured activity of $16\pm 4$ Bq/kg from $^{210}Pb$. The electrolytic copper of the refrigerator thermal shields provides an additional shield with a minimum thickness of $2$ cm. An external $10$ cm layer of borated polyethylene was installed to reduce the background due to environmental neutrons.
 
\begin{figure}[htb]
\begin{center}
\epsfig{file=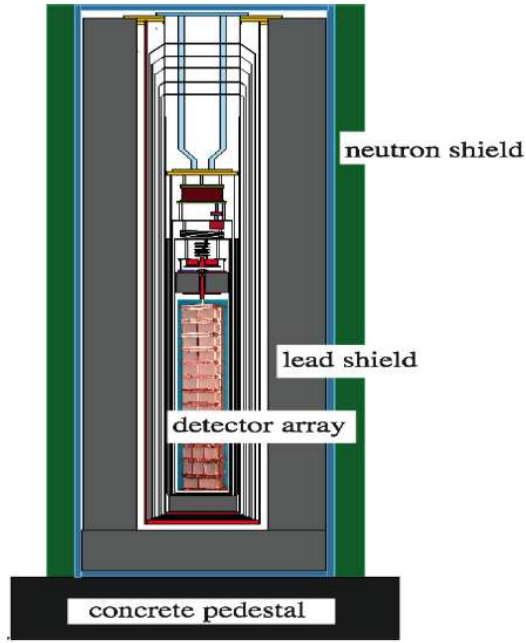, scale=0.6}
\caption{(Color online) A sketch of the CUORICINO assembly showing the tower hanging from the mixing chamber and the various heat shields and the external shielding.}
\label{fig:2}
\end{center}
\end{figure}

The detector is shielded against the intrinsic radioactive contamination of the dilution unit materials by an internal layer of $10$ cm of Roman lead ($^{210}Pb$ activity $<4$ mBq/kg \cite{50}), located inside of the cryostat immediately above the tower of the array. The background from the activity in the lateral thermal shields of the dilution refrigerator is reduced by a lateral internal shield of Roman lead that is $1.2$ cm thick. The refrigerator is surrounded by a Plexiglas anti-radon box flushed with clean $N_{2}$ from a liquid nitrogen evaporator and is also enclosed in a Faraday cage to eliminate electromagnetic interference. A sketch of the assembly is shown in Fig. \ref{fig:2}.

When cooled to $8$ mK there is a temperature spread of $\sim 1$ mK among the different detectors. Routine calibrations are performed using two wires of thoriated tungsten inserted inside the external lead shield in immediate contact with the outer vacuum chamber (OVC) of the dilution refrigerator. Calibrations normally last one to two days, and are performed at the beginning and end of each run, which lasts for approximately two-three weeks.

\begin{figure}[htb]
\begin{center}
\epsfig{file=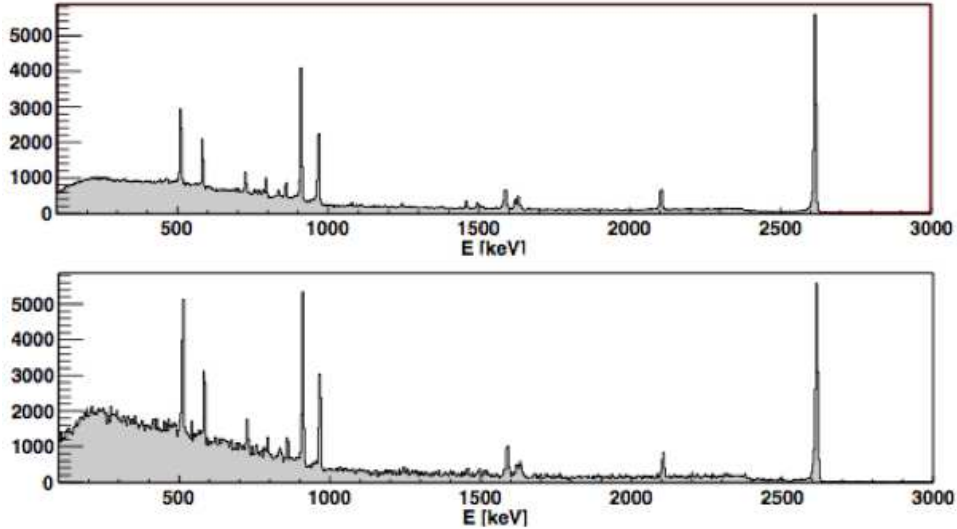, scale=0.6}
\caption{Typical calibration spectra of the CUORICINO array with a $^{232}Th$ source: $5\times 5\times 5\textrm{ cm}^{3}$ crystals upper frame, $3\times 3\times 6\textrm{ cm}^{3}$ crystals lower frame.}
\label{fig:3}
\end{center}
\end{figure}

The CUORICINO array was first cooled down at the beginning of $2003$. However, during this operation electrical connections were lost to $12$ of the $44$ detectors of $5\times 5\times 5\textrm{ cm}^{3}$, and to one of the 3$\times 3\times 6\textrm{ cm}^{3}$ crystals. Thermal stresses broke the electrical connections on their thermalizer stages that allow the transition in temperature of the electric signals in several steps from the detectors at $\sim 8$ mK to room temperature. When the cause of the disconnection was found, new thermalizer stages were fabricated and tested at low temperature. However, since the performance of the remaining detectors was normal, and their total mass was $\sim 30$ kg, warming of the array and rewiring were postponed for several months while $\nbb$-decay data were collected. At the end of $2003$, CUORICINO data acquisition was stopped and the system was warmed to room temperature and the broken thermalizer stages were replaced with new ones. During this operation, the tower was kept enclosed in its copper box to prevent possible recontamination of the detectors. As a consequence, two detectors whose disconnections were inside the box were not recovered. The same was true for one of the small central detectors whose $Si$ resistor was electrically disconnected inside the box. In the middle of $2004$, CUORICINO was cooled down and data collection began again. Typical calibration spectra are shown in Fig. \ref{fig:3}.

\section{DATA ACQUISITION AND ANALYSIS}\label{sec:5}

The signals coming from each bolometer are amplified and filtered with a six-pole Bessel low-pass filter and fed to a $16$-bit ADC. The signal is digitized with a sampling time of $8$ ms, and a circular buffer is filled. With each trigger pulse, a set of $512$ samples is recorded to disk; accordingly, the entire pulse shape is stored for offline analysis. Each channel (bolometer) has a ompletely independent trigger and trigger threshold, optimized according to the bolometerÕs typical noise and pulse shape. Starting with run No. $2$, the CUORICINO data acquisition (DAQ) now has a software trigger that  implements a "debounce" algorithm to reduce spurious fast signal triggering. The trigger is ready again within a few tens of ms, a delay due to the debounce time. Therefore, most of the pile-up events are re-triggered. The trigger efficiency above $100$ keV was evaluated as $99\pm1\%$ by checking the fraction of recorded pulser signals. The offline analysis uses an Optimal Filter technique \cite{44} to evaluate the pulse amplitudes and to compare pulse-shapes with detector response function. Events not caused by interactions in the crystals are recognized and rejected on the basis of this comparison. Pile-up pulses are identified and dealt with. This is important for calibration and high rate measurements because the pulses have long time durations and pile-up pulses can significantly increase the dead time. However, the pile-up fraction during the search for $\nbb$-decay is negligible given the low trigger rate from signals above threshold. The pile-up probability on the rise time is $\sim 0.01\%$, while that on the entire sampling window is quite a bit higher, $\sim 0.4\%$. However, these events are easily identified and the pile-up pulses are rejected. The total trigger rate, before any pulse-shape rejection, is time and channel dependent. On a single channel it ranges from a few mHz to hundreds of mHz, with a mean value of about $20$ mHz. Accepted-pulse amplitudes are then corrected using the variation in the gain measured with the heat pulses from the $Si$ resistors.  Finally, spectra are produced for each detector.

Any type of coincidence cut can be applied to the data written to disk, before the creation of the final spectra, depending on the specific analysis desired. In the case of $\beta\beta$-decay analyses, anticoincidence spectra are used. This allows the rejection of background counts from gamma rays that Compton scatter in more than one bolometer, for example. The probability of accidental coincidences over the entire detector is negligible ($<0.6\%$). Crosstalk pulses have been observed between a few channels; however, the resulting pulses are rejected on the basis of pulse-shape.


\section{SOURCE CALIBRATION AND DETECTOR\\ PERFORMANCE}\label{sec:6}

The performance of each detector is periodically checked during the routine calibration with the $^{232}Th$ gamma rays from thoriated calibration wires. The most intense gamma ray peaks visible in the calibration spectra are used. They are the: $511,583,911,968,1588$ and $2615$ keV $\gamma$-rays, and the single escape peak of the $2615$ keV gamma ray at $2104$ keV. The resulting amplitude-energy relationship is obtained from the calibration data, and the pulse amplitudes are converted into energies. The dependence of the amplitude on energy is fit with a second order log-polynomial for which the parameters were obtained from the calibration data. The selection of the functional form was established by means of simulation studies based on a thermal model of the detectors. The details of how the thermal model was applied have been published elsewhere \cite{44}. These calibration data are also used to determine the energy resolution of each bolometer. Data sets are collected for two to six weeks, separated by radioactive-source calibrations. The data collected by a single detector in this short time does not have the statistical significance to show the background gamma-ray lines because of the very low counting rates. The energy resolution, and the stability of the energy calibration, relies on the heater pulses, and on the initial and final source calibration measurements.

Double-beta decay data collected with each detector during a single data collection period are rejected if any of the following criteria are not fulfilled:

\begin{enumerate}
\item[(i)] The position of the $2615$ keV background $\gamma$-ray line from the decay of $^{208}Tl$, in the initial and the final source-calibration measurements must be stable to within $1/3$ of the measured full width at half maximum (FWHM) of the $2615$ keV line for that detector.
\item[(ii)] The energy resolution of the $2615$ keV $\gamma$-ray lines in the initial and final energy calibration measurements must be stable within $30\%$.
\item[(iii)] The energy position of the heater pulses during the entire data collection period for that data set must be stable to within $1/3$ of the characteristic (FWHM) for that detector.
\item[(iv)] The energy resolution measured with the heater pulses for that entire data collection period must be stable within $30\%$ over the entire data collection period.
\end{enumerate}

Whenever one or more of these criteria is not fulfilled, the data from that detector are not included in the final data set. Approximately $17\%$ of the data were discarded because they failed one or more of these criteria. Frequent causes of  failure to satisfy all of the criteria were noise pulses that degrade the energy resolution and temperature drifts that change the operating parameters of the bolometers. The particular bolometers involved cary; however, some are more sensitive to noise and temperature changes than others. The application of coincidence cuts does not change the efficiency; however, the difference in rise time between pulses from various bolometers can cause coincidences not to be recognized as such, cut this effect is small. in any case, the only result of the failure to recognize coincidences is the loss of background reduction, which would tend to make the quoted bound conservative. 

In both runs, the measured detector performances appear to be excellent; the average FWHM resolutions in the energy region around $2530$ keV during the calibration measurements are $7$ and $9$ keV, for the $5\times 5\times 5\textrm{ cm}^{3}$ and $3\times 3\times 6\textrm{ cm}^{3}$ detectors, respectively. The spread in the FWHM is about $2$ keV in both cases. The smaller detectors have somewhat worse resolution on average, while they also exhibit a very important nonlinearity. When the calibration spectra from all of the larger and smaller detectors are summed together, the summed spectrum resembled that of a single large detector as shown in Figure \ref{fig:3}.

\begin{figure}[htb]
\begin{center}
\epsfig{file=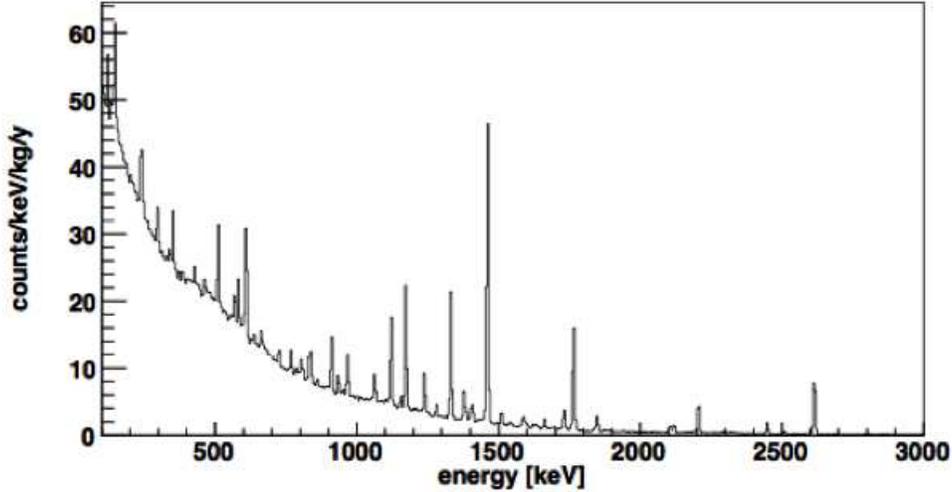, scale=0.6}
\caption{The sum spectrum of the background from the $5\times 5 \times 5\textrm{ cm}^{3}$ detectors, from both runs, to search for $\nbb$-decay.}
\label{fig:4}
\end{center}
\end{figure}


\section{DOUBLE-BETA DECAY RESULTS}\label{sec:7}

Following the shutdown discussed earlier, and restart in May $2004$, a second interruption was required to remove the malfunctioning helium liquefier used to automatically refill the main bath of the dilution refrigerator. There were also short interruptions for routine maintenance of the $17$-year old refrigerator. Excluding these interruptions, the duty cycle was very satisfactory, $\sim 60\%$ , not withstanding the fact that $15$ to $20\%$ of the live time is necessary for calibration.

The three spectra corresponding to large ($5\times 5\times 5\textrm{ cm}^{3}$) detectors and the smaller natural and enriched ($3\times 3\times 6\textrm{ cm}^{3}$) detectors are kept separate because of the different detection efficiencies for $\beta\beta$-decay events, and also because of their different background counting rates. For similar reasons, the spectra of the two runs are treated separately. Because the background rates in the spectra of Runs I and II do not show any statistically significant difference, it was concluded that no recontamination of the detector took place when the cryostat was opened to air during the interruption between Runs I and II. The full data set used in this analysis has a total effective exposure of $11.83\textrm{ kg}\cdot\textrm{yr}$ of $^{130}Te$ for the entire array.

The full summed spectrum, shown in Fig. \ref{fig:4}, clearly exhibits the $\gamma$-ray line from the decay of $^{40}K$, and those from the $^{238}U$ and $^{232}Th$ chains. Also visible are the lines of $^{57}Co$, $^{60}Co$, and $^{54}Mn$, due to the cosmogenic activation of the tellurium and the copper frame. The correct positions and widths of the peaks in the sum spectrum demonstrate the effectiveness of the calibration and linearity of the spectra. The accuracy of calibration in the $\nbb$-decay region was evaluated to be about $\pm 0.4$ keV. The details of the gamma-ray  background resulting from a preliminary analysis of Run 2 are given in Tables \ref{tab:2}, \ref{tab:3} and \ref{tab:4}. There is also clear evidence of alpha backgrounds at energies above the $2614.5$ keV gamma ray in the decay of $^{232}Tl$. A detailed analysis attributes the dominant background in the region of interest to degraded alpha particles on the surface of the copper frames. A major effort is underway to reduce this to a minimum.

\begin{table}[htp]
\caption{Gamma rays from the decay of $^{232}Th$ observed in Run II.}
\centering
\begin{minipage}{2.5in}
\centering
\vspace{5pt}
\begin{tabular}{ccccc}
\hline\hline
\\\multicolumn{1}{c}{Energy (keV)}&&\multicolumn{1}{c}{Isotope}&&\multicolumn{1}{c}{Counts/1000 h} \\
\\
\hline
$238.6$\footnote{Contains a contribution from the $U$ chain.}&&$^{212}Pb$&&$6.84\pm 0.43$\\
$338.2$&&$^{228}Ac$&&$0.89\pm 0.40$\\
$463.0$\footnote{Contains a contribution from $^{125}Sb$.}&&$^{228}Ac$&&$1.33\pm 0.25$\\
$510.7$\footnote{Contains a contribution from annihilation radiation.}&&$^{208}Tl$&&$7.78\pm 0.38$\\
$583.2$&&$^{208}Tl$&&$3.88\pm 0.30$\\
$727.3$&&$^{212}Bi$&&$1.04\pm 0.21$\\
$785.4$\footnote{Contains a contribution from $^{214}Bi$ in the $U$ chain.}&&$^{212}Bi$&&$1.02\pm 0.20$\\
$794.9$&&$^{228}Ac$&&$0.70\pm 0.25$\\
$833.0$\footnote{Contains a contribution from $^{54}Mn$.}&&$^{228}Ac$&&$2.85\pm 0.25$\\
$911.2$&&$^{228}Ac$&&$4.69\pm 0.26$\\
$964.8$&&$^{228}Ac$&&$1.37\pm 0.19$\\
$968.9$&&$^{228}Ac$&&$2.79\pm 0.21$\\
$1588.1$&&$^{228}Ac$&&$0.65\pm 0.12$\\
$1593.0$\footnote{Contains a contribution from $^{214}Bi$ in the $U$ chain.}&&$^{208}Tl$&&$0.25\pm 0.10$\\
$1620.6$&&$^{212}Bi$&&$0.58\pm 0.15$\\
$1631.0$&&$^{228}Ac$&&$0.39\pm 0.13$\\
$2614.5$&&$^{208}Tl$&&$6.90\pm 0.26$\\
\hline\hline\\
\end{tabular}
\end{minipage}
\label{tab:2}
\end{table}

\begin{table}[htp]
\caption{Gamma rays from the $^{238}U$ chain in data of Run II. Most of the activity is attributed o a radon contamination due to temporary leak in the anti-radon box surrounding the refrigerator.}
\begin{minipage}{\textwidth}
\vspace{5pt}
\begin{tabular}{ccccrcccccr}
\hline\hline
\\
\multicolumn{1}{c}{Energy (keV)}&&\multicolumn{1}{c}{Isotope}&&\multicolumn{1}{c}{Rate Cts/1000 h}&&\multicolumn{1}{c}{Energy (keV)}&&\multicolumn{1}{c}{Isotope}&&\multicolumn{1}{c}{Rate Cts/1000 h}\\
\\
\hline
$241.9$\footnote{Contains a contribution from $^{214}Pb$ in the $Th$ chain.}&&$^{214}Pb$&&$6.84\pm 0.43$&&$1401.7$&&$^{214}Bi$&&$1.23\pm 0.13$\\
$295.2$&&$^{214}Pb$&&$2.69\pm 0.48$&&$1408.0$&&$^{214}Bi$&&$1.85\pm 0.15$\\
$352.0$&&$^{214}Pb$&&$3.88\pm 0.42$&&$1509.5$&&$^{214}Bi$&&$1.85\pm 0.13$\\
$609.4$&&$^{214}Bi$&&$13.09\pm 0.47$&&$1583.2$&&$^{214}Bi$&&$0.99\pm 0.15$\\
$665.6$&&$^{214}Bi$&&$2.54\pm 0.33$&&$1594.7$\footnote{Contains a contribution from $^{208}Tl$ in the $Th$ chain.}&&$^{214}Bi$&&$0.25\pm 0.10$\\ 
$768.4$&&$^{214}Bi$&&$2.55\pm 0.33$&&$1599.3$&&$^{214}Bi$&&$0.43\pm 0.90$\\
$786.0$\footnote{Contains a contributions from $^{214}Bi$ in the $Th$ chain.}&&$^{214}Bi$&&$1.02\pm 0.20$&&$1661.5$&&$^{214}Bi$&&$1.06\pm 0.13$\\
$803.0$&&$^{210}Po$&&$1.52\pm 0.19$&&$1729.9$&&$^{214}Bi$&&$2.51\pm 0.14$\\
$934.1$&&$^{214}Bi$&&$1.75\pm 0.17$&&$1764.7$&&$^{214}Bi$&&$14.28\pm 0.38$\\
$1120.4$&&$^{214}Bi$&&$10.84\pm 0.40$&&$1838.4$&&$^{214}Bi$&&$0.40\pm 0.07$\\
$1155.3$&&$^{214}Bi$&&$1.38\pm 0.14$&&$1847.7$&&$^{214}Bi$&&$1.98\pm0.17$\\
$1238.2$&&$^{214}Bi$&&$4.83\pm 0.21$&&$2118.9$&&$^{214}Bi$&&$1.21\pm 0.12$\\
$1281.1$&&$^{214}Bi$&&$1.32\pm 0.13$&&$2204.5$&&$^{214}Bi$&&$4.55\pm 0.24$\\
$1377.8$&&$^{214}Bi$&&$3.37\pm 0.17$&&$2448.0$&&$^{214}Bi$&&$1.51\pm 0.14$\\
$1385.3$&&$^{214}Bi$&&$0.88\pm 0.88$&&&&&&\\
\hline\hline\\
\end{tabular}
\end{minipage}
\label{tab:3}
\end{table}

\begin{table}[htp]
\caption{Background gamma rays from a variety of sources including isotopes produced by cosmogenic neutrons: $^{60}Co$, $^{54}Mn$, and fall out isotopes $^{137}Cs$, $^{207}Bi$.}
\begin{minipage}{\textwidth}
\vspace{5pt}
\begin{tabular}{ccccrcccccr}
\hline\hline
\\
\multicolumn{1}{c}{Energy (keV)}&&\multicolumn{1}{c}{Isotope}&&\multicolumn{1}{c}{Rate Cts/1000 h}&&\multicolumn{1}{c}{Energy (keV)}&&\multicolumn{1}{c}{Isotope}&&\multicolumn{1}{c}{Rate Cts/1000 h}\\
\\
\hline
$122.1$&&$^{57}Co$&&$5.39\pm 0.44$&&$661.7$&&$^{137}Cs$&&$1.26\pm 0.19$\\
$427.9$&&$^{125}Sb$&&$1.95\pm 0.27$&&$834.8$\footnote{Contains a contribution from $^{228}Ac$ in the $Th$ chain.}&&$^{54}Mn$&&$2.86\pm 0.25$\\
$463.2$\footnote{Contains a contribution from $^{228}Ac$ in the $Th$ chain.}&&$^{125}Sb$&&$1.33\pm 0.25$&&$1063.7$&&$^{207}Bi$&&$2.36\pm 0.29$\\
$511.0$\footnote{Contains a contribution from $^{208}Tl$ in the $Th$ chain.}&&annihilation&&$7.78\pm 0.38$&&$1173.2$&&$^{60}Co$&&$11.6\pm 0.33$\\
$569.7$&&$^{207}B$&&$3.11\pm 0.27$&&$1332.5$&&$^{60}Co$&&$11.9\pm 0.36$\\
$600.6$&&$^{125}Sb$&&$1.42\pm 0.20$&&$1461.0$&&$^{40}K$&&$31.4\pm 0.58$\\
$635.9$&&$^{125}Sb$&&$0.64\pm 0.18$&&$2505.7$&&$^{60}Co$&&$0.31\pm 0.05$\\
\hline\hline\\
\end{tabular}
\end{minipage}
\label{tab:4}
\end{table}
 
\begin{figure}[htb]
\begin{center}
\epsfig{file=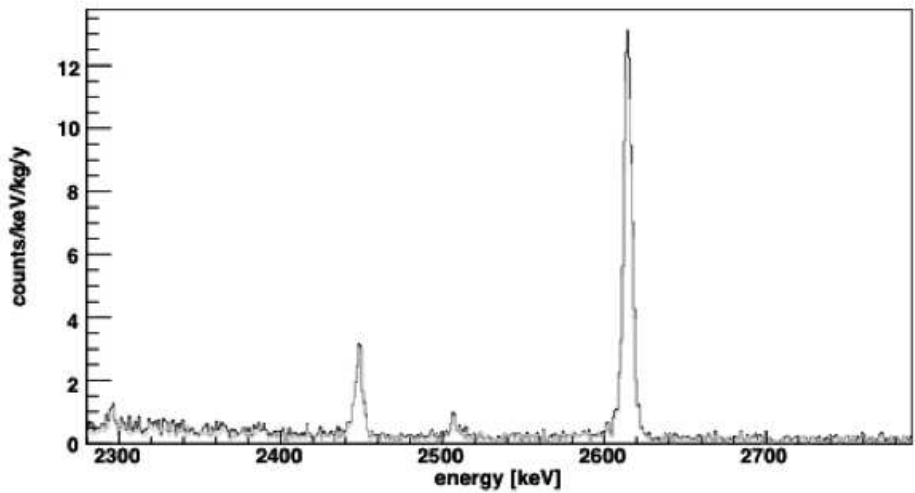, scale=0.6}
\caption{The summed background spectrum in the $\sim 400$ keV region of interest, which includes the $\nbb$-decay energy $2530.3\pm 2.0$ keV.}
\label{fig:5}
\end{center}
\end{figure}

The average background counting rates in the region of $\nbb$ decay are: $0.18\pm 0.01$, and $0.20\pm 0.04$ counts per keV, per kg, per year ($\textrm{keV}^{-1}\textrm{kg}^{-1}\textrm{y}^{-1}$) for the $5\times 5\times 5\textrm{ cm}^{3}$ and $3\times 3\times 6\textrm{ cm}^{3}$ crystals, respectively. The sum background spectrum from about $2300$ to $2700$ keV, of the $5\times 5\times 5\textrm{ cm}^{3}$ and $3\times 3\times 6\textrm{ cm}^{3}$ crystals, is shown in Fig. \ref{fig:5}. The shape of the background in the region of interest does not change when anticoincidence requirement is applied. An extensive analysis of the background contributions implies that the continuum background in the region of interest around $2530$ keV breaks down as follows: $10\pm 5\%$ is due to surface contamination of the $TeO_2$ crystals with $^{238}U$ and $^{232}Th$; $50\pm 20\%$ is due to surface contamination of the copper surfaces facing the bolometers also with $^{232}Th$ and $^{238}U$; and $30\pm 10\%$ is due to the tail of the $2614.5$ keV gamma ray in the decay of $^{232}Th$ from the contamination of the cryostat copper shields. Finally, there were no observable gamma-ray lines associated with neutron interactions. Monte-Carlo simulations of the neutron shield imply that the background from neutron interactions would be negligible.
 
\begin{figure}[htb]
\begin{center}
\epsfig{file=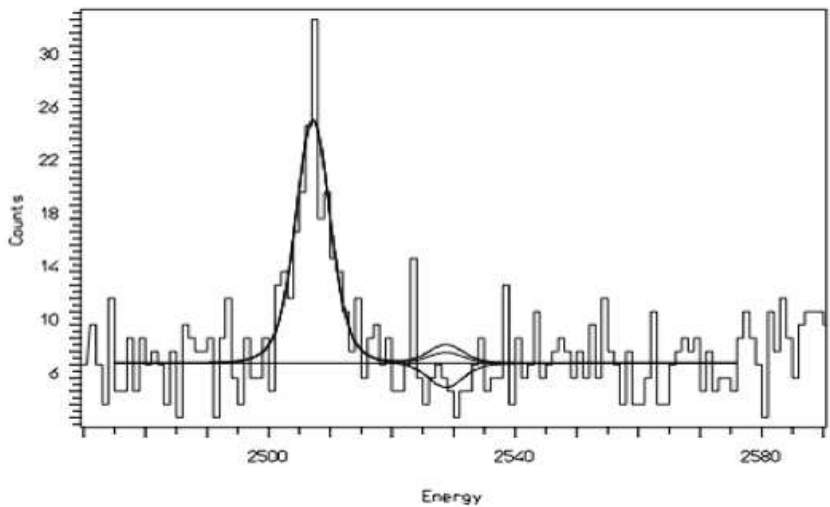, scale=0.6}
\caption{The total background spectrum from $2470$ to $2590$ keV. Clearly visible is the sum peak at $2505.74$ keV due to the sum of the $1173.24$ and $1332.50$ keV $\gamma$-ray cascade in the decay of $^{60}Co$. This activity is attributed to the $^{60}Co$ in the copper frames generated by cosmic ray neutrons while the frames were above ground. The solid lines are the best fit to the region using polynomials of the order $0$ to $2$. The three lines in the region of interest are for bounds ($68\%$ and $90\%$) CL on the number of candidate $\beta\beta$-decay events.}
\label{fig:6}
\end{center}
\end{figure}

The energy resolution for the complete data set was computed from the FWHM of the $2615$ keV background $\gamma$-ray line in the decay of $^{203}Tl$ at the end of the thorium chain. The results are $8$ keV for the forty operating $5\times 5\times 5\textrm{ cm}^{3}$ crystals, and $12$ keV for the eighteen $3\times 3\times 6\textrm{ cm}^{3}$ crystals. Clearly visible is the peak at about $2505$ keV due the summing of the $1332.50$-$1173.24$ keV $\gamma$-ray cascade in the decay of $^{60}Co$. This is $25.46$ keV, i.e., about $7$ sigma of the Gaussian energy resolution peak from the $\nbb$-decay end-point energy of $^{130}Te$, and could make a negligible contribution to the region under the expected $\nbb$-decay peak .  The sum spectrum from $2290$ to $2700$ keV is shown in Fig. \ref{fig:5}. The sum spectrum from $2470$ to $2590$ keV is shown in Fig. \ref{fig:6}.
 
The details of the operating conditions and parameters of the two CUORICINO data collection periods are given in Table \ref{tab:5}. The total usable exposure for Run I + Run II is $11.83\textrm{ kg}\cdot\textrm{yr}$ of $^{130}Te$. The event detection efficiencies were computed with Monte-Carlo simulations; they are $0.863$ and $0.845$ for the large and small crystals, respectively. The loss of efficiency of the bolometers is due to beta particles created near the surface that escape part of their energy. From the above exposure data we compute: $\ln{2}\times N_{L}\times\epsilon_{L}\times t=2.809\times10^{25}$ yr, for the large and $\ln{2}\times N_{S}\times\epsilon_{S}\times t=4.584\times10^{24}$ yr for the small crystals. Here, $\epsilon$ is the detection efficiency, while $N_{L}$ and $N_{S}$ are the numbers of $^{130}Te$ nuclei in the large and small detectors, respectively.

\begin{table}[htdp]
\caption{Summary of operating parameters for the two CUORICINO data collection periods. From columns $1$ through $8$ are listed: the run number, number of large and small detectors, the active mass of $^{130}Te$, total run time, the calibration time, the time collecting  $\bb$ - decay data, the total exposure in $\textrm{kg}\cdot\textrm{yr}$, and the usable exposure in $\textrm{kg}\cdot\textrm{yr}$ after rejection of data not fulfilling the quality requirements. The total usable exposure is then $11.83\textrm{ kg}\cdot\textrm{yr}$.}
\vspace{5pt}
\begin{center}
\begin{tabular}{cccccccc}
\hline\hline
\\
Run $\#$& Detectors & Active mass & Run time & Calibration & t-$\bb$ & Collected & Used \\
& large/small & [kg $^{130}$Te] & [d] & [d] & [d] & [kg$\cdot$yr $^{130}$Te] & [kg$\cdot$yr $^{130}$Te]  \\
\\
\hline
1 &  29/15 & 7.95 & 240 & 24.5 & 55.08 & 1.2 & 1.06 \\
2 & 40/15 & 10.37 & 983 & 108.5 & 415.1 & 11.79 & 10.77 \\
\hline\hline

\end{tabular}
\end{center}
\label{tab:5}
\end{table}

The $\beta\beta$-decay half-life limit was evaluated using a Bayesian approach. The peaks and continuum in the region of the spectrum centered on the $\beta\beta$-decay energy were fit using a maximum likelihood analysis \cite{57,58}. The likelihood functions of six spectra (the sum spectra of the three types of crystals in the two runs) were combined allowing for a different background level for each spectrum, and a different intensity of the $2505$ keV $^{60}Co$ sum peak. Other free parameters are the position of the $^{60}Co$ peak and the number of counts under a peak at the $\beta\beta$-decay energy. The same procedure is used to evaluate the $90\%$ CL limit to the number of counts present in the $\nbb$-decay peak.

Assuming Poisson statistics for the binned data, the fit procedure was formulated in terms of the likelihood chi-square analysis as described in the following equation: 

\begin{equation*}
\chi_{L}^{2}=2\sum_{j=1}^{6}\sum\left(y_{i,j}-n_{i,j}+n_{i,j}\ln{\left(n_{i,j}/y_{i,j}\right)}\right),
\end{equation*} 

where $j$ indicates the $j^{th}$ spectrum, $n_{i,j}$ is the  number of events in the $i^{th}$ bin of the $j^{th}$ spectrum and $y_{i,j}$ is the number of events predicted by the fit model. 
 
Fit parameters were estimated minimizing the $\chi_{L}^{2}$, while limits were obtained, after proper renormalization, considering the $\chi_{L}^{2}$ distribution in the physical region. The response function for each spectrum is assumed to be a sum of symmetric gaussian functions, each having the typical energy resolution of one of the detectors summed in that spectrum. The experimental uncertainty in the transition energy is considered by means of a quadratic (gaussian) term in the above equation. In the region between $2575$ and $2665$ keV, assuming a flat background, the best fit yields a negative number of counts under the peak ($-13.9\pm 8.7$). However, the resulting upper bound on the number of candidate events in the $\nbb$-decay peak is  $n=10.7$ at $90\%$ C.L. These values are normalized to a hypothetical sum spectrum of the entire statistical data set in which each of the six spectra are weighted according to the corresponding exposure, geometric efficiency, and isotopic abundance. The resulting lower limit on the half-life is computed as:
 
\begin{align*}
T_{1/2}^{0\nu}\left(^{130}Te\right)&\geq\ln2{\left\{N_{L}\epsilon_{L}+N_{S}\epsilon_{S}\right\}t/n\left(90\%\textrm{ CL}\right)}\\
&=\left(3.27\times10^{25}/10.7\right)\textrm{ yr}=3.0\times 10^{24}\textrm{ yr}.
\end{align*}

The dependence of the value of the limit on systematic uncertainties that arise from the method of analyzing the data was investigated in detail. These uncertainties reside in the dead time, energy calibration, $Q$-value, and background spectral shape. The main factor influencing the limit is the uncertainty in the background spectral shape. 

For example, changing the degree of the polynomial used to fit the background in the $\nbb$-decay region from $0$ to $2$ as well as the selection of the energy window used in the analysis can vary the bound from $2.5$ to $3.3\times10^{24}$ yr. The quoted $90\%$ CL lower bound was computed using the central value, $2530.3$ keV of the measured double beta decay energy \cite{49}. There is a small dip in the data centered at $\sim 2530$ keV as shown in Figure \ref{fig:5}. This has been treated as a statistical fluctuation. 
    
\section{NUCLEAR STRUCTURE ISSUES }\label{sec:8}

There is one theoretical viewpoint that holds that the required model space for $^{130}Te$ is still very large for reliable shell model calculations and must be severely truncated. Accordingly, the quasiparticle random-phase approximation (QRPA) is commonly used \cite{59,60,61,62,63,64,65,66,67,68,69,70,71,72,73,74,75,76,77,78}. The results from these calculations, from author to author had, until recently, differed significantly for the same nucleus. In Table \ref{tab:6}, only the results from references \cite{62,73} differ significantly from the other $13$; they correspond to the largest matrix elements.  In the QRPA approach, the particle-particle interaction is fixed by a parameter, $g_{pp}$, which is derived in various ways by different authors. Two recent papers by Rodin, \emph{et al.}, give detailed assessments of the uncertainties in QRPA calculations of $\nbb$-decay matrix elements, and explain many of the reasons for the disagreements between the various authors over the years \cite{60,61}. The numerical values given in these articles were corrected in a later erratum \cite{78}.  In Table \ref{tab:6} we list the values of  $\left\langle m_{\nu}\right\rangle$ corresponding to $T_{1/2}^{0\nu}\left(^{130}Te\right)\geq3.0\times10^{24}$ yr derived using the calculations of various authors. More details are discussed later, including the results from recent shell model calculations.

\begin{table}[htdp]
\caption{Various values of $\left\langle m_{\nu}\right\rangle$ corresponding to $T_{1/2}^{0\nu}\left(^{130}Te\right)=3.0\times10^{24}$ yr.} 
\vspace{5pt}
\begin{center}
\begin{tabular}{lcc}

\hline\hline
\\
Authors/Reference & Method & $\left\langle m_{\nu}\right\rangle$ (eV) \\
\\
\hline
\cite{78} Rodin \it{et al.}, 2007 & using $2\nu\bb$-decay to fix $g_{pp}$ & 0.46 \\
\cite{62} Staudt \it{et al.}, 1992 & pairing (Bohm) & 0.19 \\
\cite{63} Pantis \it{et al.}, 1996 & no \it{p-n} pairing & 0.52 \\
\cite{64} Vogel, 1986 & & 0.47 \\
\cite{65} Civitarese and Suhoen 2006 & & 0.42 \\
\cite{66} Tomoda, 1991 & & 0.42 \\
\cite{67} Barbero, \it{et al.}, 1999 & & 0.33 \\
\cite{68} Simkovic, 1999 & \it{pn} - RQRPA & 0.68 \\
\cite{69} Suhoen \it{et al.}, 1992 & & 0.64 \\
\cite{67} Muto \it{et al.}, 1989 & & 0.39 \\
\cite{71} Stoica \it{et al.}, 2001 & & 0.60 \\
\cite{72} Faessler \it{et al.}, 1998 & & 0.55 \\
\cite{73} Engel \it{et al.}, 1989 & seniority & 0.29 \\
\cite{74} Aunola \it{et al.}, 1998 & & 0.41 \\
\cite{79} Caurier \it{et al.}, 2008 & Nuclear Shell Model & 0.58 \\
\hline\hline
\end{tabular}
\end{center}
\label{tab:6}
\end{table}

Extracting the effective Majorana mass of the electron neutrino from the half-life requires the calculation of the nuclear structure factor, $F_{N}\equiv G^{0\nu}\left(M_{F}^{0\nu}-\left(g_{A}/g_{F}\right)^{2}M_{GT}^{0\nu}\right)$, in Eq. \eqref{eq:7}. This is not straightforward for the nuclei that are the best candidates for $\nbb$-decay experiments, e.g., $^{130}Te$, because they have many valence nucleons. To create a tractable shell-model calculation for these heavy nuclei, it is necessary to truncate the model space to the point that could affect the reliability of the results. Accordingly, schematic models are employed. As stated above, QRPA has become the standard approach for both $2\nu\beta\beta$ and $\nbb$ decay. The results calculated with QRPA, however, depend on the selection of a number of parameters, and the fact that different authors select the parameters in various ways has resulted in large differences in the resulting matrix elements as discussed in Ref. \cite{61}.
	
In Table \ref{tab:6}, we list $14$ different values of  $\left\langle m_{\nu}\right\rangle$  derived with QRPA and with renormalized QRPA, (RQRPA), corresponding to $T_{1/2}^{0\nu}\left(^{130}Te\right)=3.0\times10^{24}$ yr, and also the recent shell-model calculations of Caurier \emph{et al.} \cite{79}. From the table it is clear that the different ways of applying the same basic model has lead to a spread in the resulting matrix elements, and hence in the corresponding value of $\left\langle m_{\nu}\right\rangle$, of a factor of three \cite{61,62,63,64,65,66,67,68,69,70,71,72,73,74}. This corresponds to differences of a factor of nine in the predicted half-life for a given value of $\left\langle m_{\nu}\right\rangle$, if all calculations are given the same weight. This assumption, however, cannot be justified. It should be recognized that calculation techniques, as well as computational power have made significant progress over the years, improving the reliability of both QRPA and shell-model calculations.
 
In their recent article, Rodin, Simkovic, Faessler, and Vogel (\textit{T\"{u}bingen}) \cite{61}, give detailed discussions of how the choices of various parameters in similar models can lead to such discrepancies. These are the gap of the pairing interactions, the use of (renormalized) RQRPA that partially accounts for the violation of the Pauli principle in the evaluation of the two-fermion commutators, the nucleon-nucleon interaction potential, the strength of the particle-hole interactions of the core polarization, the size of the model space, and the strength of the particle-particle interaction, parameterized by the quantity $g_{pp}$. The matrix elements of the virtual transitions through states with $J^{\pi}=1^{+}$ in the intermediate nucleus are extremely sensitive to the value of $g_{pp}$, which makes $2\nu\beta\beta$-decay matrix elements also very sensitive to it because this decay mode only proceeds through $1^{+}$ intermediate states. On the other hand, $\nbb$-decay also proceeds via higher multipoles through states of higher spin. These transitions are found to be far less sensitive to the value of $g_{pp}$. For this reason, Rodin \emph{et al.} select the value of $g_{pp}$ that makes the calculation of the $2\nu\beta\beta$-decay half-life agree with the experimental value. In addition, some calculations are greatly simplified by using an average energy in the denominator of the second-order matrix-element expression, and the sum over the intermediate states is done by closure. When the value, $g_{A}=1.245$, of the axial-vector coupling constant obtained from muon decay is used, it commonly lead to a value of the Gamow-Teller strength typically larger than the measured value. To ameliorate this situation, a quenched value $g_{A}=1.00$ is used. In calculated rates of $2\nu\beta\beta$-decay, which proceed only through $J^{\pi}=1^{+}$ states, this results in a factor of $2.44$ reduction in the rate. Using the technique of Rodin \emph{et al.} \cite{61}, the choice of $g_{aA}=1.00$ reduces the rate by between $10$ to $30\%$, depending on the particular nucleus.

Another serious difference between some of the $\nbb$-decay calculations is due to the treatment of the short-range correlations in the nucleon-nucleon interactions. It was also pointed out by Simkovic \emph{et al.} \cite{68}, that including the momentum dependent higher order terms of the nucleon current typically result in a reduction in the calculated value of the $\nbb$-decay matrix element by about $30\%$. These were included in the calculations of Refs. \cite{60,61}.
	
In recent paper by Alvarez \emph{et al.} \cite{75}, a QRPA formalism for $2\nu\beta\beta$-decay in deformed nuclei was presented. A considerable reduction in the matrix elements was observed in cases in which there was a significant difference in the deformations of the parent and daughter nuclides. Exactly how this would affect $\nbb$-decay calculations is not yet clear. It must be understood that this uncertainty, when resolved could result in a further reduction in neutrinoless double-beta decay matrix elements calculated within the framework of QRPA and RQRPA.

In general, however, the paper by Rodin \emph{et al.} \cite{61}, represents a detailed study of the various factors that cause the large variations in the nuclear matrix elements of $\nbb$-decay calculated by different authors over the years, and must be taken seriously. The procedure of Rodin \emph{et al.} \cite{59,60,61} has the attractive feature that it gives a straightforward prescription for selecting the very important particle-particle parameter, $g_{pp}$. However, Civitarese and Suhonen (referred to as the \textit{Jyv\"{a}skyl\"{a}} group) have given strong arguments in favor of using single $\beta^{\pm}$-decay and electron capture data for this purpose, while giving arguments against using experimental $2\nu\beta\beta$-decay half lives \cite{65}. They argue that only states with spin and parity $1^{+}$ can be the intermediate states involved in $2\nu\beta\beta$-decay, and that in the neutrinoless process these states play a minor role, and that the higher spin states play a dominant role. The \textit{Jyv\"{a}skyl\"{a}} group recently presented a preprint in which they show that the effects of short-range correlations have been significantly overestimated in the past \cite{76,77}. Accordingly, their matrix elements originally gave a very different picture of the of the physics impact of the CUORICINO data presented in this paper. However, recently there have been some very important developments discussed below.


\section{RECENT DEVELOPMENTS IN\\ QRPA CALCULATIONS}\label{sec:9}

We adopt the position that the large dispersion in values in the nuclear matrix elements implied by the values in Table \ref{tab:6} does not reflect the true state of the art. Instead, we assume that there has been significant progress in understanding the key theoretical issues, as well as large increases in available computational power. Until very recently, however, two of the recent extensive theoretical treatments of the $\nbb$-decay matrix elements disagreed significantly, and in particular in the case of $^{130}Te$. The relevant nuclear structure factors, $F_{N}$, from the \textit{Jyv\"{a}skyl\"{a}} and \textit{T\"{u}bingen} groups for $g_{A}=1.25$ were $F_{N}\left(^{130}Te\right)=1.20\pm 0.27\times 10^{-13}\textrm{ yr}^{-1}$ of Rodin \emph{et al.} \cite{61}, and $F_{N}\left(^{130}Te\right)=5.13\times 10^{-13}\textrm{ yr}^{-1}$ of Civitarese and Suhonen \cite{65}.

Recently an erratum was submitted by Rodin \emph{et al.} \cite{78} with major corrections to Table \ref{tab:1} of Ref. \cite{61}. A coding error was discovered in the computation of the short-range correlations that, for example, increased the predicted $\nbb$-decay rate of $^{130}Te$ by a factor of $4.03$. Their corrected value of the nuclear structure factor of $^{130}Te$, is now $F_{N}\left(^{130}Te\right)=4.84^{+1.30}_{-0.64}\times 10^{-13}\textrm{ yr}^{-1}$, in good agreement with the above value given by Civitarese and Suhonen. However, there is still a small disagreement between these two groups concerning the technique for calculating short-range correlations. Rodin \emph{et al.}, used a Jastrow-correlation function, which has subsequently been shown by Kortelainen \emph{et al.} \cite{76} to overestimate the effects of short-range correlations, and hence to result in an excessive reduction in the nuclear matrix elements.

Kortelainen \emph{et al.} \cite{77} have also updated the calculations of Civitarese and Suhonen. They extended their model space, for the cases of $^{116}Cd$,  $^{128,130}Te$ and $^{136}Xe$, to include the $1p$-$0f$-$2s$-$1d$-$0g$-$2p$-$1f$-$0h$ single particle orbitals, calculated with a spherical Coulomb-corrected Woods-Saxon potential. In Ref. \cite{77}, a complete discussion is given of their method of fixing the parameters of the Hamiltonian. In this treatment they fix particle-particle parameter $g_{pp}$ of the pnQRPA using the method of Rodin \emph{et al.} \cite{59,60,61}, namely with the experimentally measured $2\nu\beta\beta$-decay half-lives. They did not use the Jastrow-correlation function to correct for the short-range correlations, but rather they employ a "unitary correlation operator method" (UCOM), which in the case of $^{130}Te$ increases the matrix element by a factor of $1.38$ over that calculated with the Jastrow correlation function. Their new values for the nuclear structure factors are:

\begin{align*}
F_{N}\left(^{130}Te\right)_{g_{A}=1.25}&=7.47\times10^{-13}\textrm{ yr}^{-1}, \\
F_{N}\left(^{130}Te\right)_{g_{A}=1.00}&=4.93\times10^{-13}\textrm{ yr}^{-1}.
\end{align*}

This is to be compared to the results of the earlier work of Civitarese and Suhonen \cite{65}.

In any case, the major disagreements between the \textit{Jyv\"{a}skyl\"{a}} and \textit{T\"{u}bingen} groups have finally been understood, and the present difference in the predicted $\nbb$-decay rates of $^{130}Te$ now differ by a factor of $1.06$, whereas the earlier disagreement was by a factor of $4.28$. Some remaining differences might well lie in the differing methods of applying the short-range correlations (see also the discussion in Ref. \cite{80}). In any case these recent developments have had a major impact on the interpretation of the CUORICINO data.
 
Furthermore, the group of Caurier \emph{et al.} \cite{79}, have recently given new values for these matrix elements from improved nuclear shell model calculations. The shell-model matrix elements are somewhat smaller than those of the recent \textit{Jyv\"{a}skyl\"{a}} and corrected \textit{T\"{u}bingen} results, and according to their matrix elements, the CUORICINO data imply: $\left\langle m_{\nu}\right\rangle\leq 0.58$ eV.


\section{CUORICINO AS A TEST OF THE CLAIM\\ OF DISCOVERY}\label{sec:10}

The CUORICINO array is the only operating $\nbb$-decay experiment, with energy resolution adequate to potentially probe the range of effective Majorana mass, $\left\langle m_{\nu}\right\rangle$, implied by the observation of $\nbb$-decay claimed by Klapdor-Kleingrothaus \emph{et al.} \cite{19,20}. In the $2006$ article by Klapdor-Kleingrothaus and Krivosheina (KK\& K) \cite{20}, the peak in the spectrum centered at $Q_{\beta\beta}\cong 2039$ keV is interpreted as the $\nbb$-decay of $^{76}Ge$, consistent with the range: $T_{1/2}^{0\nu}\left(^{76}Ge\right)=\left\{1.30-3.55\right\}\times10^{25}\textrm{ yr }(3\sigma)$. The best-fit value is $\left(2.23_{-0.31}^{+0.44}\right)\times10^{25}$ yr. In this discussion we offer no critique of the claim, however, since this claim has been criticized from several points of view \cite{21,22,23}, it is interesting to ask if it is feasible to observe a $\nbb$-decay with this half-life with a significant confidence level with the published parameters of the experiment. Below, we show that the answer is "yes", the experiment could have made the observation in the range of half-lives quoted \cite{20}.
	
It is straightforward to derive an approximate analytical expression for the half-life sensitivity for discovery at a given confidence level that an experiment can achieve (see Appendix). The achievable discovery half-life, when the background rate is nonzero, is expressed as:

\begin{equation}
\label{eq:9}
T_{1/2}^{0\nu}\left(n_{\sigma}\right)=\frac{4.17\times10^{26}\textrm{ yr}}{n_{\sigma}}\left(\frac{\epsilon a}{W}\right)\sqrt{\frac{Mt}{\left(1+\zeta\right)b\,\delta(E)}}.
\end{equation}

It is more conventional to simply have $b\,\delta(E)$ in the denominator of the root of Eq. \eqref{eq:9} as prescribed by the Particle Data Book \cite{81}. However, when the background continuum is obtained by a best fit to all peaks and continuum in the region, we choose this alternative approach. In Eq. \eqref{eq:9}, $n_{\sigma}$ is the desired number of standard deviations of the (CL) ($3$ for $\textrm{CL}=99.73\%$, for example), $\epsilon$ is the event detection and identification efficiency, $a$ is the isotopic abundance, $W$ is the molecular weight of the source material, $M$ is the total mass of the source, $\zeta$ is the signal-to-background ratio, $b$, is the specific background rate in counts/keV/kg/yr, and $\delta(E)$ is the instrumental width of the region of interest related to the energy resolution at the energy of the expected $\nbb$-decay peak.

The values for these parameters for the Heidelberg-Moscow experiment \cite{17,19,20} are:  $Mt=71.7\textrm{ kg}\cdot\textrm{yr}$, $b=0.11\textrm{ kg}^{-1}\textrm{keV}^{-1}\textrm{yr}^{-1}$, $\epsilon=0.95$, $a=0.86$, $W=76$, and $\delta(E)=3.27\textrm{ keV}$. The number of counts under the identified peak at $2039$ keV is $28.75\pm 6.86$. The average value of the background near the region of interest was $11.6$ counts, therefore $\zeta\cong 2$. Direct substitution into Eq. \eqref{eq:9} yields:

\begin{equation}
\label{eq:10a}
T_{1/2}^{0\nu}\left(4\sigma,\,^{76}Ge\right)=0.9\times 10^{25}\textrm{ yr};\quad T_{1/2}^{0\nu}\left(3\sigma\right)=1.2\times 10^{25}\textrm{ yr}.
\end{equation}

Using the less conservative approach with $b\,\delta(E)$ in the denominator, the predicted half-life sensitivity for a discovery is

\begin{equation}
\label{eq:10b}
T_{1/2}^{0\nu}\left(4\sigma,\,^{76}Ge\right)=1.6\times 10^{25}\textrm{ yr};\quad T_{1/2}^{0\nu}\left(3\sigma\right)=2.13\times 10^{25}\textrm{ yr}.
\end{equation}

These are close to the claimed most probable value given in Ref. \cite{20}. This simple analysis is independent of the claimed result, with the exception of the determination of the signal to background ratio, $\zeta$. The conclusion is that with the given experimental parameters, this experiment could have had a discovery potential. Since this analysis does not account for statistical fluctuations, the discovery confidence level could possibly fall between $3\sigma$ and $5\sigma$. Any criticism of the claim would involve a reanalysis of the data, and the interpretation of the background peaks in the region. This falls outside of the scope of this discussion. Accordingly, we do not question the claim, but rather ask how well the present CUORICINO data confront it, now and in the future after five years of running.

While the many theoretical calculations of the nuclear matrix elements over the years have differed significantly, the recently corrected-QRPA calculations of \emph{T\"ubingen} \cite{78},  those of \emph{Jyv\" askyl\" a} \cite{65}, and shell model calculations of Caurier \emph{et al.} \cite{79}, differ by less than about $30\%$. We have chosen to use for further analysis of the physics impact of the present CUORICINO data.

Equation \eqref{eq:8} can be inverted to obtain the values of the nuclear structure factor, $F_{N}$, using the calculated half-lives for $\nbb$-decay calculated with a given $\left\langle m_{\nu}\right\rangle$ by the authors of the theoretical papers. The resulting values are as follows:

$^{76}Ge_{g_{A}=1.245}$:
\begin{align}
\label{eq:13}
&\textrm{Rodin, \emph{et al.}:}&F_{N}&=1.22^{+0.10}_{-0.11}\times10^{-13}\textrm{ yr}^{-1},\nonumber \\
&\textrm{Caurier, \emph{et al.}:}&F_{N}&=4.29\times10^{-14}\textrm{ yr}^{-1},\\
&\textrm{Civitarese and Suhonen:}&F_{N}&=7.01\times10^{-14}\textrm{ yr}^{-1}\nonumber
\end{align}

$^{130}Te_{g_{A}=1.245}$:
\begin{align}
\label{eq:14}
&\textrm{Rodin, \emph{et al.}:}&F_{N}&=4.84^{+1.30}_{-0.64}\times10^{-13}\textrm{ yr}^{-1}\textrm{ (corrected value)},\nonumber \\
&\textrm{Caurier, \emph{et al.}:}&F_{N}&=2.57\times10^{-13}\textrm{ yr}^{-1},\\
&\textrm{Civitarese and Suhonen:}&F_{N}&=5.13\times10^{-13}\textrm{ yr}^{-1}.\nonumber
\end{align}

The resulting values and ranges of values of $\left\langle m_{\nu}\right\rangle$ implied by the KK\&K data, and by the CUORICINO data are as follows:

\begin{align}
\label{eq:15}
&\left\langle m_{\nu}\right\rangle_{\textrm{KK\&K}}^{\textrm{Rod}}=\{0.23-0.43\}\textrm{ eV},\nonumber \\
&\left\langle m_{\nu}\right\rangle_{\textrm{CUOR}}^{\textrm{Rod}}\leq\{0.38-0.46\}\textrm{ eV},\nonumber \\
&\left\langle m_{\nu}\right\rangle_{\textrm{KK\&K}}^{\textrm{Civ}}=\{0.32-0.54\}\textrm{ eV},\\
&\left\langle m_{\nu}\right\rangle_{\textrm{CUOR}}^{\textrm{Civ}}\leq0.41\textrm{ eV},\nonumber\\
&\left\langle m_{\nu}\right\rangle_{\textrm{KK\&K}}^{\textrm{Cau}}=\{0.41-0.68\}\textrm{ eV},\nonumber \\
&\left\langle m_{\nu}\right\rangle_{\textrm{CUOR}}^{\textrm{Cau}}\leq0.58\textrm{ eV}.\nonumber
\end{align}

The results of the analyses with the new corrected matrix elements of Ref. \cite{78} imply that the CUORICINO sensitivity has entered well into the range of values of $\left\langle m_{\nu}\right\rangle$ implied by the claim of KK\&K. In the other two analyses, the CUORICINO data also constrain part of the range of values of $\left\langle m_{\nu}\right\rangle$ implied by KK\&K.\\

It is also interesting to try to predict the sensitivity of CUORICINO if it were to continue to operate for a total of $5$ years. The three recent calculations of the nuclear matrix elements result in the following predicted decay rates if the Heidelberg claim is correct. In this case, the decay rates would be:

\begin{align}
\label{eq:16}
\tau_{\textrm{KK\&K}}^{-1}\left(^{76}Ge\right)&=\{1.95-5.32\}\times10^{-26}\textrm{ yr}^{-1},\nonumber \\
\tau_{\textrm{Rod}}^{-1}\left(^{130}Te\right)&=\{0.62-2.94\}\times10^{-25}\textrm{ yr}^{-1}, \\
\tau_{\textrm{Civ}}^{-1}\left(^{130}Te\right)&=\{1.43-3.89\}\times10^{-25}\textrm{ yr}^{-1},\nonumber\\
\tau_{\textrm{Cau}}^{-1}\left(^{130}Te\right)&=\{1.17-3.19\}\times10^{-25}\textrm{ yr}^{-1}.\nonumber
\end{align}

Accordingly, we can calculate the number of $\nbb$-decay counts with $5$ years of live-time operation expected in the CUORICINO data consistent with the claim of KK\&K. The exposure would be: $Nt\epsilon=2.85\times10^{26}$ y, resulting in the following predicted number of real $\nbb$-decay events:

\begin{align}
\label{eq:17}
\tau_{\textrm{Rod}}^{-1}Nt\epsilon&=\{18-84\}_{\nbb},\nonumber \\
\tau_{\textrm{Civ}}^{-1}Nt\epsilon&=\{41-110\} _{\nbb},\\
\tau_{\textrm{Cau}}^{-1}Nt\epsilon&=\{33-91\}_{\nbb}.\nonumber
\end{align}

These counts would be superimposed on an expected background of $35$ to $39$ counts per keV in the $8$ keV region of interest centered at $2530$ keV. 

The constraints placed by the current CUORICINO data might favor the lower numbers in the ranges above. This would make it more challenging for CUORICINO to confirm the discovery claim of KK\&K, and renders it almost impossible to rule out the KK\&K claim with a significant level of confidence. The solution to this problem is the construction and operation of the proposed first tower of CUORE, called CUORE-$0$, combine its data with  that of CUORICINO, and later the complete CUORE Experiment. 


\section{THE PROPOSED CUORE EXPERIMENT}\label{sec:11}

The proposed CUORE detector will be made of $19$ towers of $TeO_{2}$ bolometers, very similar to the CUORICINO tower \cite{28}. Each will house $13$ modules of four $5\times 5\times 5\textrm{ cm}^{3}$ crystals with masses of $\sim 750$ g. CUORE will contain $\sim 200$ kg of $^{130}Te$. The $988$ bolometers will have a total detector mass of $\sim 750$ kg and will operate at $8$-$10$ mK. An intense research and development program is underway to reduce the background to $0.01$ counts/keV/kg/yr. Thus far a reduction has been achieved that has reached within a factor of $2.4$ of this goal in the region of $2530$ keV, the $Q$-value for the $\nbb$-decay of $^{130}Te$. With this background, CUORE would reach a sensitivity of $\sim T_{1/2}^{0\nu}\left(^{130}Te\right)\approx 2.1\times10^{26}$ yr in $5$ years. The secondary goal is to achieve a background level of $0.001$ counts/keV/kg/yr. This would allow a half-life sensitivity of $T_{1/2}^{0\nu}\approx 6.5\times10^{26}$ yr. 

In case that the background would be reduced to $0.001$ counts/keV/kg/yr, the associated sensitivities in the effective Majorana mass of the electron neutrino, $\left\langle m_{\nu}\right\rangle$, would be

\begin{align}
\label{eq:18}
\left\langle m_{\nu}\right\rangle_{\textrm{Rod}}&=\{0.026-0.031\}\textrm{ eV},\nonumber \\
\left\langle m_{\nu}\right\rangle_{\textrm{Civ}}&=0.028\textrm{ eV},\\
\left\langle m_{\nu}\right\rangle_{\textrm{Cau}}&=0.040\textrm{ eV}.\nonumber
\end{align}

The half-life sensitivity is directly proportional to the abundance, $a$, of the parent $\beta\beta$-decay isotope [see equation \eqref{eq:9}]. Accordingly, enriching the detectors of CUORE from $33.8\%$ in $^{130}Te$ to $90\%$, CUORE would achieve the same sensitivity with a background of $0.01$ counts/keV/kg/yr as it would with natural $Te$ and a background of $0.0014$ counts/keV/kg/yr. An R\&D program, to determine the feasibility and cost of isotopically enriching CUORE is underway . In addition, the CUORE collaboration has a rigorous R\&D program to improve the energy resolution from an average of $8$ keV, as it is in CUORICINO, to $5$ keV. This resolution should be achievable because some of the CUORICINO bolometers have already achieved $5$ keV. An intense program is underway to determine the cause of the spread in energy resolution. If in the end, CUORE does achieve the background of $0.001$ counts/keV/kg/yr, in addition is enriched, and has an average energy resolution of $5$ keV, it could reach a half life sensitivity of $2.5\times 10^{27}$ yr in $10$ years. 

In this case the sensitivities become:

\begin{align}
\label{eq:19}
\left\langle m_{\nu}\right\rangle_{\textrm{Rod}}&=\{13-16\}\textrm{ meV},\nonumber \\
\left\langle m_{\nu}\right\rangle_{\textrm{Civ}}&=14\textrm{ meV},\\
\left\langle m_{\nu}\right\rangle_{\textrm{Cau}}&=20\textrm{ meV}.\nonumber
\end{align}

This brings the sensitivity into the normal hierarchy region, which exceeds the goals of some of the other next generation experiments. It is possible to proceed as planned with a natural abundance version of CUORE, and then the bolometers could be replaced with those isotopically enriched in $^{130}Te$. This would increase the half-life reach by a factor of $2.5$ for an enrichment of $85\%$.  


\section{SUMMARY AND CONCLUSIONS}\label{sec:12}

The CUORICINO detector is an array of $62$ $TeO_{2}$ bolometers operating at a temperature of about $8$ mK. It has a total mass of $40.7$ kg of $TeO_{2}$, containing $11$ kg of $^{130}Te$. It has operated for a total exposure of $N\left(^{130}Te\right)t\epsilon=5.47\times10^{25}$ yr, with no observation of $\nbb$-decay events, results in a lower bound, $T_{1/2}^{0\nu}\left(^{130}Te\right)\geq 3.0\times 10^{24}$ yr. The corresponding upper bound on the effective Majorana mass of the electron neutrino, $\left\langle m_{\nu}\right\rangle$, using the corrected nuclear structure calculations of Rodin \emph{et al.}, is $\left\langle m_{\nu}\right\rangle\leq(0.38-0.46)$ eV, while using those of Civitarese and Suhonen yields $\left\langle m_{\nu}\right\rangle\leq 0.47$ eV. With the recent shell model calculations the CUORICINO data imply $\left\langle m_{\nu}\right\rangle\leq 0.58$ eV. In all cases, the present CUORICINO data probe a significant portion of the range of the half life measured by KK\&K. If the Heidelberg claim is correct, the nuclear structure calculations of Ref. \cite{78} imply that after $5$ years of live time CUORICINO would detect $\{18-84\}$, $\nbb$-decay events, while those of Ref. \cite{65} imply it would detect $\{41-110\}$ events, and those of Ref. \cite{79} imply it would detect $\{33-91\}$ $\nbb$-events. In all cases, these counts would appear in Gaussian peaks with $\textrm{FWHM}=8k$ keV, superimposed on an average background of $35-39$ counts keV$^{-1}$.

In any case, the current results imply that the continued operation of CUORICINO is very important since it represents the only possibility of testing the claim of evidence of $\nbb$-decay for the next $5$ years or more.


\section*{ACKNOWLEDGEMENTS}
The CUORICINO Collaboration owes many thanks to the Directors and Staff of the Laboratori Nazionali del Gran Sasso over the years of the development, construction and operation of CUORICINO, and to the technical staffs of our Laboratories. The experiment was supported by the Istituto Nazionale di Fisica Nucleare (INFN), the Commission of the European Community under Contract No. HPRN-CT-$2002$-$00322$, by the U.S. Department of Energy under Contract No. DE-AC$03$-$76$-SF$00098$, and DOE W-$7405$-Eng-$48$, and by the National Science Foundation Grants Nos. PHY-$0139294$ and PHY-$0500337$. We also wish to thank the following colleagues for their help and advice: Juoni Suhonen, Osvaldo Civiterese, Petr Vogel, Amand Faessler. Vadim Rodin, and Fedor Simkovic, and Fernando Ferroni.

\appendix
\section{APPENDIX}

\begin{figure}[htb]
\begin{center}
\epsfig{file=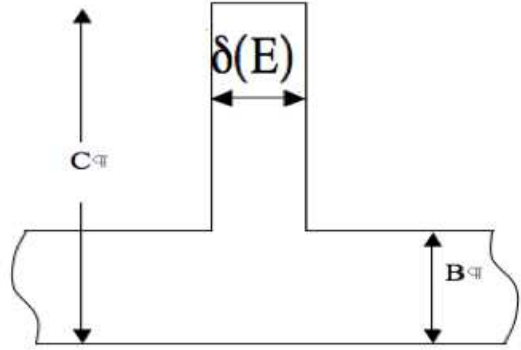, scale=0.6}
\caption{Diagram showing the scheme on which Eq. \eqref{eq:a2} is derived.}
\label{fig:7}
\end{center}
\end{figure}

An approximate expression for estimating the $\nbb$-decay half-life at which a given experiment can achieve discovery at the confidence level corresponding to $n_{\sigma}\sigma$, can be derived by reference to Figure \ref{fig:7}. Let $"C"$ be the total number of counts found in the region of the expected $\nbb$-decay peak; let $"B"$ be the total number of background counts in the same energy interval, $\delta(E)$. For the number of real $\nbb$-decay events to have a statistical significance of $n_{\sigma}$, the following must be true: $C-B=n_{\sigma}\sqrt{C}$. In the usual case where $B\neq 0$, a desired signal to background ratio, $\zeta\equiv(C-B)/B$, can be chosen; hence $C=\left(1+\zeta\right)B$. The usual expression for the corresponding half-life can be written in terms of these parameters as:

\begin{equation}
\label{eq:a1}\tag{A.1}
T_{1/2}^{0\nu}\left(n_{\sigma}\right)=\frac{(\ln{2})Nt\epsilon}{n_{\sigma}\sqrt{\left(1+\zeta\right)B}}
\end{equation}

where $N$ is the total number of parent nuclei, $\epsilon$ is the total detection efficiency, and $t$ is the live time of the data collection. The number of parent nuclei can be written in terms of, $M$, the total mass of the source (in an oxide for example), as follows: $N=\left(10^{3}\textrm{ g/kg/Wq/mole}\right)\cdot\left(A_{0}\textrm{ at/mole}\right)\cdot a\left(\textrm{abundance}\right)\cdot M\textrm{kg}$. Substituting these values, and expressing the background in terms of the background rate, $B=bM\delta(E)t$, where $b=\textrm{(counts/keV/kg/yr)}$, the expression is written:

\begin{equation}
\label{eq:a2}\tag{A.2}
T_{1/2}^{0\nu}\left(n_{\sigma}\right)=\frac{4.17\times 10^{26}}{n_{\sigma}}\left(\frac{a\epsilon}{W}\right)\sqrt{\frac{Mt}{\left(1+\zeta\right)b\delta(E)}}
\end{equation}

Of course in the case of zero background, Eq. \eqref{eq:a1} is used, and the quantity, $\left(1+\zeta\right)B$, is replaced the number of real events in the peak. In case there are no real or background events, i.e., $C=B=0$, the denominator of Eq. \eqref{eq:a1} is replaced by the usual quantity, $\ln{\{1/(1-CL)\}}$, which is $2.3$, ($90\%$ C.L.)  for example, and $T_{1/2}^{0\nu}$ becomes an experimental lower limit. In Eq. \eqref{eq:a2}, we use the fluctuation in the real events instead of that of the background because in these experiments the background level used is that of a best fit curve to the background in the region, and the fluctuation is a fitting error and is much smaller than the statistical fluctuations in the region of interest.


\bibliographystyle{plain}

\end{document}